\documentclass{iopart}
\usepackage{graphicx}
\begin{document}
\title{Scatter Networks: A New Approach for Analyzing Information Scatter}

\author{
Lada A. Adamic$^1$, Suresh K. Bhavnani$^1$ and Xiaolin Shi$^2$}
\address{1. School of Information, University of Michigan, Ann Arbor, MI 48109}
\address{2. Electrical Engineering and Computer Science, University
of Michigan, Ann Arbor, MI 48109} \ead{ladamic@umich.edu}
 \maketitle
\begin{abstract}
Information on any given topic is often scattered across the Web.
Previously this scatter has been characterized through the
inequality of distribution of facts (i.e. pieces of information)
across webpages. Such an approach conceals how specific facts (e.g.
rare facts) occur in specific types of pages (e.g. fact-rich pages).
To reveal such regularities, we construct bipartite networks,
consisting of two types of vertices: the facts contained in webpages
and the webpages themselves. Such a representation enables the
application of a series of network analysis techniques, revealing
structural features such as connectivity, robustness, and
clustering. Not only does network analysis yield new insights into
information scatter, but we also illustrate the benefit of applying
new and existing analysis techniques directly to a bipartite network
as opposed to its one-mode projection. We discuss the implications
of each network feature to the users' ability to find comprehensive
information online. Finally, we compare the bipartite graph
structure of webpages and facts with the hyperlink structure between
the webpages.
\end{abstract}

\section{Introduction}

Information on any given topic tends to be scattered on the Web. No
one page seems to have all the facts, and no one fact seems to be in
all the relevant pages.  But besides this simple observation of the
fact frequency, very little is understood about how the information
is actually distributed. In this work, we extend prior notions of
information scatter by constructing a bipartite graph of documents
and the facts (pieces of information) they contain. The networks we
construct are not the typical networks of hyperlinked webpages.
Rather, pages are linked together by shared information content and
information is linked to other information by co-occurrence in
webpages. The notion of webpages being linked by the common facts
they contain is not as far from users' search behavior as it may
first seem.  A Web user seeking comprehensive information might
explore a new topic by alternately querying a search engine with
keywords, reading a webpage and modifying the search queries to
incorporate the new information learned. Therefore, how a user
explores a topic space while using a search engine is directed both
by this bipartite network of webpages and facts and by the
hyperlinks being traced. While much research has gone into
developing algorithms for ranking search engine results using
hyperlinks and semantic content, little has been done to explore how
comprehensive the information available to the user is. Our novel
approach of representing the scatter of information as a {\em
scatter network} allows for an exploratory view of this previously
obscured phenomenon.

To illustrate the utility of our approach, we apply qualitative
techniques such as visualization and quantitative network analysis
metrics to a sample network of pages about the medical condition
melanoma, and the facts about melanoma they contain. The
visualization, in addition to making apparent the highly skewed
distribution of facts, also reveals a clustering of facts and
documents into sub-topics. We use a bipartite measure of clustering
to show that the information is indeed highly clustered and a
community finding algorithm to automatically identify those
clusters. We apply proximity measures to infer how quickly a user
exploring this bipartite network would find related facts. Crucial
to rapid fact discovery are those facts and documents that bridge
multiple topics. We identify them through a network centrality
measure called betweenness~\cite{Kwang-IlGoh10012002}. To determine
just how crucial these bridging facts and documents are, we perform
a robustness analysis where we remove a fraction of the documents
and examine if the different facts are still connected through
document co-occurrence.

Given that most novice users have difficulty finding comprehensive
healthcare information on the Web~\cite{bhavnani06strategyhub}, we
discuss implications for search and retrieval in each part of our
network analysis. For example, a measure called assortativity tells
us whether facts contained in fact-rich pages tend to be common or
rare. If the latter is the case, one can simply point a user to
several fact-rich pages and expect that the information across them
should be comprehensive. If the former is the case, however, the
fact-rich pages will contain redundant information, while the rare
facts will be hidden by themselves in other documents. Finally, we
compare the scatter network structure with the hyperlink structure
in order to answer the question of how rapidly a user can access
comprehensive information. Although our present study is exploratory
in nature, our findings have important implications to the design of
both search engines and the websites they are covering.

\subsection{Background}
Several studies have analyzed the distribution of content across
information sources at different levels of granularity. These
include the distribution of articles across
journals~\cite{bradford48}, the distribution of words within a
book~\cite{zipf49human}, the distribution of articles across online
databases~\cite{tenopir82,lancaster85,Hood01}, and the distribution
of facts about a topic across webpages and
websites~\cite{bhavnani05scatter,over98}. In each case, the studies
analyzed the relationship of one variable (e.g. number of relevant
articles) against another variable (e.g. number of journals) through
a distribution analysis. Each of the resulting distributions was
highly skewed. As we have already mentioned, a few facts tend to
occur in many pages, while many facts occur in a few. Similarly, a
few pages contain many facts while most contain just a few. This
consistent result has led researchers to believe that skewed
distributions are a stable property of how information tends to
exist across information sources~\cite{bates02}, a phenomenon
commonly referred to as information scatter. In an earlier
study~\cite{bhavnani05scatter} we speculated that the amount of
detail (e.g. a single sentence versus a paragraph) devoted to a fact
on a page could explain why there are so many pages with few facts
about a common healthcare topic across high-quality healthcare
sites. Although the above studies reveal the complex and dynamic
nature of content and links on the Web, little is understood about
the regularities underlying the scatter of facts across pages.

Networks have been studied in a wide variety of fields, ranging from
sociology to biology to computer science~\cite{newman03structure}.
Network analysis has particular relevance to the structure of the
World Wide Web, a network with billions of pages that are connected
through even more numerous hyperlinks~\cite{Broder00,ladamicsw}. The
structure of the Web has implications in search engines' coverage,
as well as their ability to rank relevant
results~\cite{page98pagerank,kleinberg98authoritative}. Relatively
few studies have focused on bipartite networks directly, without
first taking their one-mode projections. But considering, for
example, a bipartite model of authors and articles, explains much of
the degree distributions and clustering coefficients observed in the
one-mode projection network of
co-authors~\cite{newman01graphs,ramasco:036106,donetti:188701}.

\begin{figure}[thbp]
\begin{center}
\includegraphics[width=1\columnwidth]{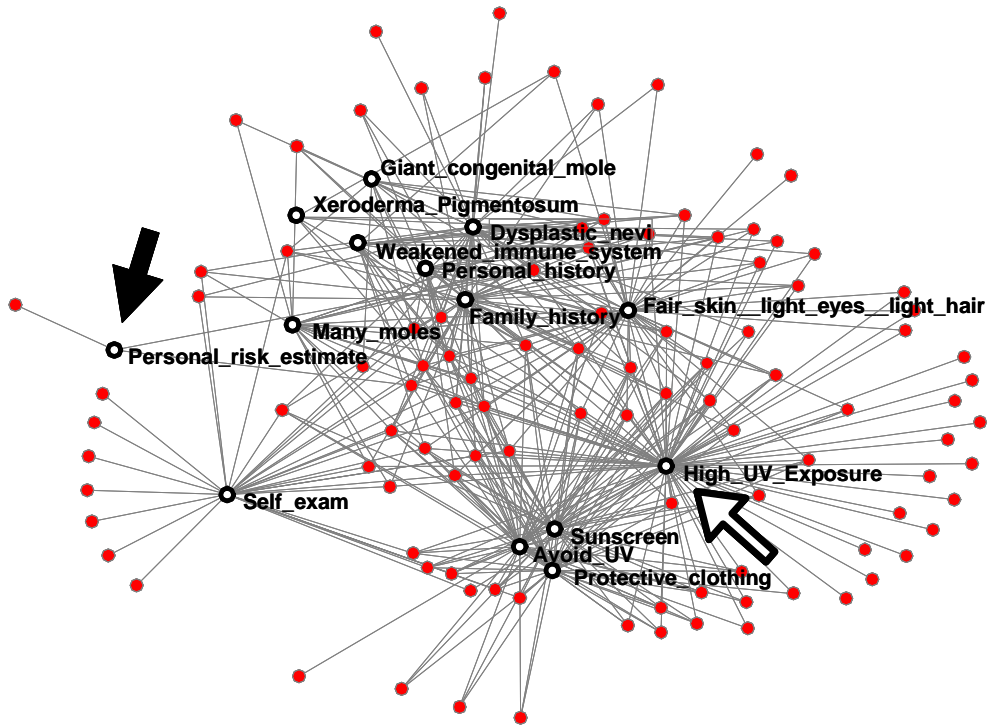}
\caption{A spring layout of the network of melanoma risk and
prevention facts (white) and the documents that contain them
(solid)~\cite{bhavnani07scatter} drawn using the GUESS network
analysis and visualization software~\cite{1124889}.\label{risknet}}
\end{center}
\end{figure}

 In applying
network analysis techniques to scatter networks, we build on
previous work on the scatter of facts across webpages, with the goal
of understanding regularities within the scatter, and the
implications of those regularities on approaches to help users find
more comprehensive information.

\section{Methodology}\label{datagathermethod}
\subsection{Description of data set}
Our data set, obtained from a previous
study~\cite{bhavnani05scatter}, consists of a set of webpages on the
topic of melanoma, and the set of facts that are contained within
them. The data was gathered in several steps:
\begin{enumerate}
\item A set of facts
(e.g. high UV exposure increases your risk of getting melanoma) were
identified about 5 melanoma topics: risk/prevention, self-exams,
diagnosis, doctor's visits, and staging. \item These facts were
verified and expanded upon by a panel of experts, in this case
doctors who are melanoma specialists.
\item A search query was generated for each fact by search experts.
\item 10 high quality websites
were selected by asking specialists and from the MedlinePlus listing
of sites containing information on
melanoma\footnote{www.nlm.nih.gov/medlineplus/melanoma.html}.
\item The Google search engine was used to retrieve webpages which
matched queries corresponding to the facts from within the 10
authoritative sites with skin cancer information.
\item Human judges evaluated whether a fact was present in each of the
webpages retrieved.
\end{enumerate}

This process resulted in 53 facts about 5 melanoma topics that
occurred in 336 relevant pages from the top-10 websites with skin
cancer information. Because natural language processing techniques
are so far unable to reliably identify facts~\cite{peck06fact}, and
human evaluation, while much more reliable, is quite slow, the data
set is necessarily relatively small. We believe, however, that our
analysis shows the applicability of network concepts to the study of
information scatter, and already provides several novel insights
even from the small subset of pages on the topic of melanoma.

\subsection{Constructing and visualizing the scatter network}
We construct our scatter network by drawing an edge between each
webpage and each fact contained in that webpage. Because edges exist
only between two {\em different} types of nodes, the facts and
pages, this is what is known as a two-mode or bipartite graph. While
we will shortly describe a number of quantifiable characteristics of
networks, much can be learned by visualization, especially if the
networks are of small to moderate size. Figure~\ref{risknet} shows a
spring layout of the network of melanoma risk facts and the
documents that contain them. The spring layout places vertices that
are connected to each other close together, as if they are connected
by springs that pull them together, while allowing vertices that are
not connected to drift apart. It is immediately apparent that three
facts placed close together near the bottom of the layout are
closely related because they co-occur in many of the same documents.
These are the three facts pertaining to preventing melanoma:
avoiding UV radiation, wearing sunscreen, and wearing protective
clothing.  In addition we see that some facts are important enough
to be mentioned by themselves in certain documents. These facts are:
self-exams can help prevent more advanced melanoma and that high UV
exposure causes melanoma. We will compare this bipartite scatter
network to the one-mode hyperlink network of webpages in
Section~\ref{hyperlink}.

\begin{figure}[htbp]
\begin{center}
\includegraphics[width=1\columnwidth]{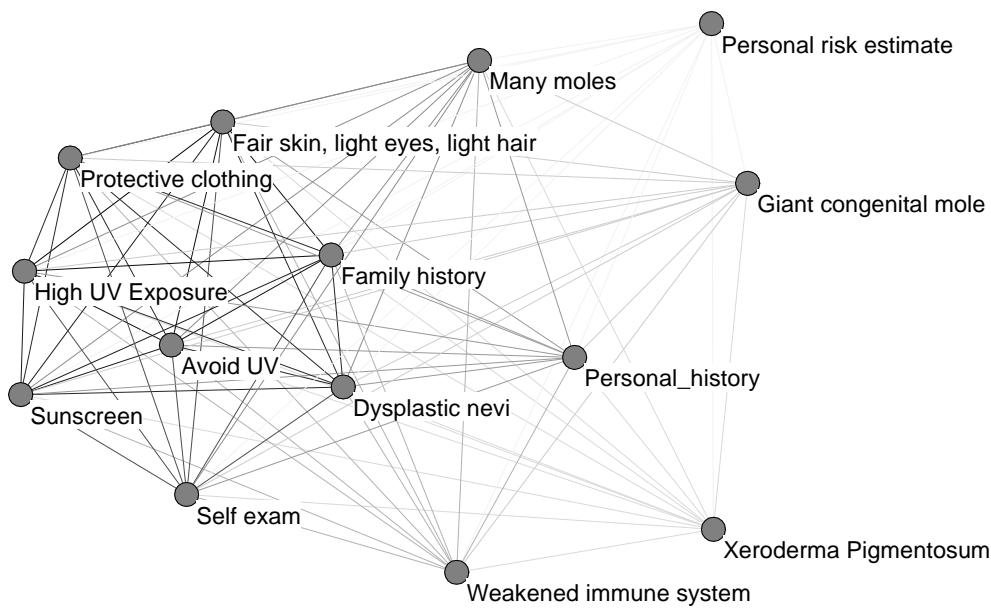}
\caption{ The one-mode fact network corresponding to a projection of
the two-mode fact and document network in Figure~\ref{risknet}. The
width and shading of the lines corresponds to the number of
documents containing both facts.\label{onemodewlabels}}
\end{center}
\end{figure}

One can transform a two mode network into a one mode network by
considering, e.g. two facts linked if they co-occur on at least one
webpage. Such a one-mode projection is shown
Figure~\ref{onemodewlabels}. Note that such a transformation is
lossy, that is we no longer know which documents the facts
co-occurred in. One can however assign weights to each edge
corresponding to the number of shared connections in the two-mode
network. Since most network metrics are designed only for unweighted
networks, and since omitting weights introduces a further loss of
information, we will prefer to work with the full bipartite network
directly.

\section{Network analysis\label{networkanalysis}}
To explore the nature of information scatter, we use a series of
network measures, most of them quite well established, but almost of
which have so far been primarily applied to one-mode networks.
Through these measures we can characterize (1) the regularities in
the occurrence of facts across pages, and (2) the connectivity
between pages and facts corresponding to different topics. Some of
these analyses adapt a particular meaning when applied to bipartite
networks, and each has implications for search and retrieval.
\subsection{Regularities in the occurrence of
facts across pages}

\subsubsection{Degree distributions} 
\begin{figure}[htbp]
\begin{center}
\includegraphics[width=0.75\columnwidth]{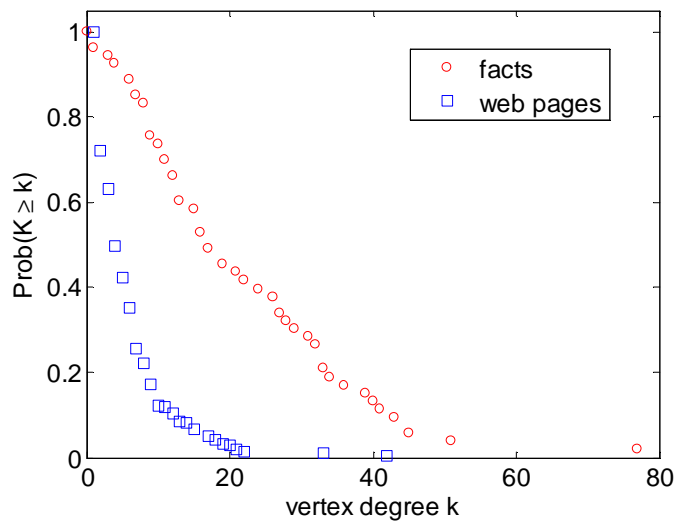}
\caption{The cumulative degree distributions of documents and facts
combined over five melanoma topics.\label{degreedist}}
\end{center}
\end{figure}

Prior
studies have determined that the distribution of facts in documents
is highly skewed. A few documents have many facts - many documents
have a few. In our bipartite network, this statement describes the
degree distribution of the document vertices. The same tends to hold
for the degree distribution of the facts. A few ``common" facts are
mentioned in many documents, but the more numerous ``rare" facts are
mentioned in just a few. As shown in Figure~\ref{degreedist}, the
degree distribution of pages is more skewed than that of facts, in
part because pages containing facts tend to be more numerous than
the facts themselves. Note that, unlike the hyperlink distribution
of the web graph~\cite{albert00scalefree,Broder00}, these
distributions are not power-law. However, this is quite possibly due
to the limitation in sampling - we do not have all the documents in
the Web that contain a fact (there are over a million pages
mentioning both UV and melanoma, but we only have a limited sample
from selected medical websites). Likewise, there are probably
hundreds of obscure facts that didn't make it on the list. In
general, a sample of a power-law network may not necessarily yield
power laws~\cite{Stumpf05}.  What matters here is that the degree
distribution confirms that many pages have few facts, while some
have many, but not all.

The skew in the distribution may be due to a form of preferential
attachment process. A page that already has many facts is more
likely to add another, since its purpose may be to provide a wide
range of information. Similarly, a fact that is contained in many
pages is more likely to be learned and re-iterated on a new or
existing page. In general, various mechanisms may separately lead to
heavy tailed distributions~\cite{newman05powerlaw}. A sketch of a
generative model particular to scatter networks is given
in~\cite{bhavnani07scatter}. This brings us to the next question of
where the rare facts reside.

\subsubsection{Degree correlations\label{assortativity}}
We next look at degree correlations in order to answer the question
of whether pages containing rare facts are fact-rich or fact poor.
We measure this with degree-degree correlations corresponding to the
Pearson correlation coefficient between the degree of the webpage
(the number of facts it contains) and the degree of the fact in the
webpage (the total number of documents that contain
it)~\cite{newman03structure}. Social networks, for example, tend to
be assortative (have positive degree correlation) - people who know
lots of people tend to know other people who know lots of people.
This is particularly true of social networks constructed from
bipartite graphs, for example the actor collaboration network or
scientific collaboration networks. A movie with a large cast will
generate a large, fully connected clique, with each member having a
large degree and sharing edges with neighbors who by definition also
have large degree. Many technological networks, on the other hand,
are disassortative. For example, in the physical internet, highly
connected regional hubs will connect lower degree vertices
~\cite{Maslov04internet,vespignaniPRLbackbone}. Biological networks
such as protein-protein interaction networks were also found to be
disassortative~\cite{SergeiMaslov05032002}. Note that in all of
these networks, one measures the assortativity among vertices of the
same type, e.g. do the co-authors of high degree authors themselves
have high degree?

\begin{figure}[htbp]
\begin{center}
\includegraphics[width=0.8\columnwidth]{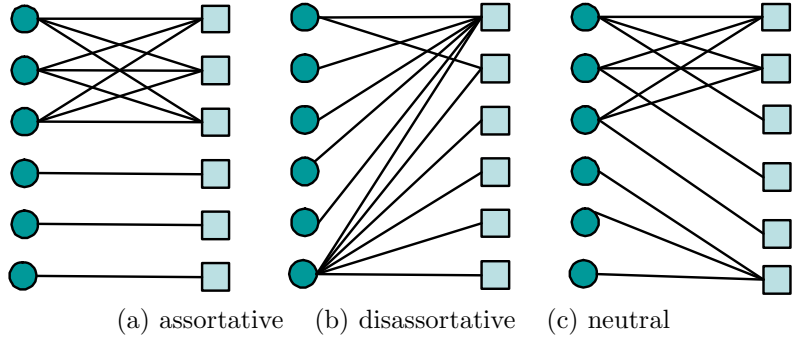}
\begin{tabular}{@{\extracolsep{\fill}}lcr}
  (a) assortative & (b) disassortative & (c) neutral\\
\end{tabular}
\caption{Three networks (circles representing facts on the left,
squares representing documents on the right) with the same numbers
of documents and facts, and edges between them, but different
assortativities.\label{assortativediagram}}
\end{center}
\end{figure}

In the context of scatter networks, it makes most sense to view
assortativity from a bipartite, and novel, perspective. Rather than
asking whether the one-mode projection of facts or webpages is
assortative, we look at the degree correlations of the webpages and
facts in relation to each other, as shown in
Figure~\ref{assortativediagram}. The diagram on the left shows a toy
assortative network (Pearson correlation $\rho = 1$), with high
degree documents connected to high degree facts, and low degree
documents linking to low degree facts. In other words, assortativity
implies that pages rich in facts will mention the most common facts
that are found in many other documents, while pages that mention
only a single fact will tend to mention less widely discussed facts.
The diagram in the middle is a disassortative network ($\rho =
-0.74$). In this example, there is a single page mentioning all the
facts, and one fact that is mentioned in all the pages, most of
which mention only that one fact. In actuality, we find that
information scatter tends to be disassortative in four of the five
topics: risk/prevention (-0.26**), doctor's exam (-0.35***),
diagnostic tests (-0.17*), and disease stages
(-0.56***)\footnote{***, **,  and
* denote significance at the 0.001, 0.01 and .05 levels
respectively.}. For example, in the risk/prevention scatter network
(Figure~\ref{risknet}), the common fact {\em High UV Radiation}
(close to the white arrow) occurs in many fact-poor pages,
presumably because of its importance to preventing melanoma.
Furthermore, the rare fact {\em Personal Risk Estimate} (close to
the black arrow) occurs in two fact-rich pages in the center of the
graph.

In contrast, for the topic Self-exam, there was a positive degree
correlation (0.14*). Here, fact-rich pages tended to contain the
same common facts related to mole appearance, and rare facts such as
resources for locating a dermatologist, were located on pages
containing few or no other facts. What can we conclude from the
above observations of negative assortativity? Primarily, while
information is scattered, it is the common facts that appear on
their own, while the rare facts tend to be covered along with many
of the other facts in comprehensive pages. Note that the
disassortativity is closely linked to the skewed degree
distributions. Because the networks are small, constraining the
degrees constrains the structural variety of randomized versions of
the graph~\cite{Holme07randomized,alderson07connectivity}. In
particular, the skewed degree distribution dictates that the
fact-rich pages are likely to contain several of the rare facts in
order to satisfy their degree constraint~\cite{Maslov04internet}. In
all subtopic networks, randomized versions of the network which
conserve degrees have negative assortativity. For example, a
randomized version of doctor's visit topic has an expected
degree-degree correlation of $-0.25\pm0.03$, slightly less
disassortative than the observed $-0.35$. Interestingly, even for
the self-exam topic, where the measured correlation is positive, the
degree correlation of a randomized network is negative
($-0.21\pm0.03$).

What implications do degree correlations have for a person seeking
health information? If they have access to a search engine that
ranks fact-rich pages highly, they will have immediate access to
many, if not most, of the facts in the first few documents they
access. For example, if information is distributed as in
Figure~\ref{assortativediagram} b) and the document with all the
facts is ranked most highly by the search engine, it will not matter
that all the other documents mention only one or two facts. However,
if information is distributed as in Figure~\ref{assortativediagram}
a), and the search engine ranks the most fact rich pages most
highly, then the user will find the exact same common facts in the
first 3 search results, and may finish her search before accessing
the specific documents that have additional facts. If on the other
hand the user is accessing documents at random, disassortativity
(Figure~\ref{assortativediagram}b) means that she will have to
access many documents mentioning the same common fact, and only with
small probability find the one with all the facts. In contrast, in
the assortative network of Figure~\ref{assortativediagram}a), even
if she accesses a document with just one fact, this fact will likely
be a new one.

\subsection{Connectivity between topics}
\subsubsection{Connected components\label{connected}}
\begin{figure}[htbp]
\centering
\includegraphics[width=0.70\columnwidth]{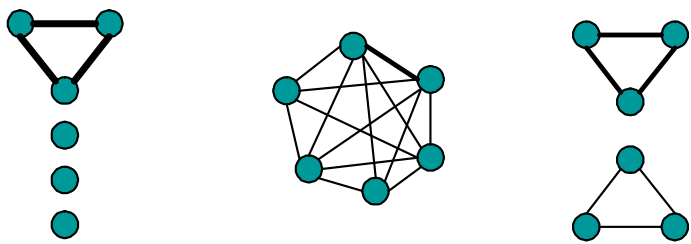}
\caption{One mode projections of the networks in
Figure~\ref{assortativediagram}, showing the connected
components.}\label{onemodediagram}
\end{figure}

When considering information scatter, we may be interested in
whether a user can discover all the facts by traversing the
bipartite network. This is not a straightforward traversal as is the
case for the Web, where the user reaches documents by following
hyperlinks. Rather, for the present we omit any consideration of
hyperlinks that may exist between the documents (after all, these
need not be hyperlinked documents, but rather documents stored in an
online database). In the bipartite network two facts are linked if
they are in the same document(s), therefore a user can learn about
one fact while reading the document with regard to another. The user
may then search for the second fact and read a different document
that may expose them to further facts, etc. This may not be the
exact mechanism by which a user discovers the facts, just as a web
surfer will not tirelessly traverse all the possible hyperlinks.
Instead, it is a conceptual question whether one could discover all
the facts by performing such a traversal. Taking the networks in
Figure~\ref{assortativediagram}, we can construct the following
1-mode networks of facts shown in Figure~\ref{onemodediagram}.

\begin{figure}[htbp]
\centering
\includegraphics[width=0.95\columnwidth]{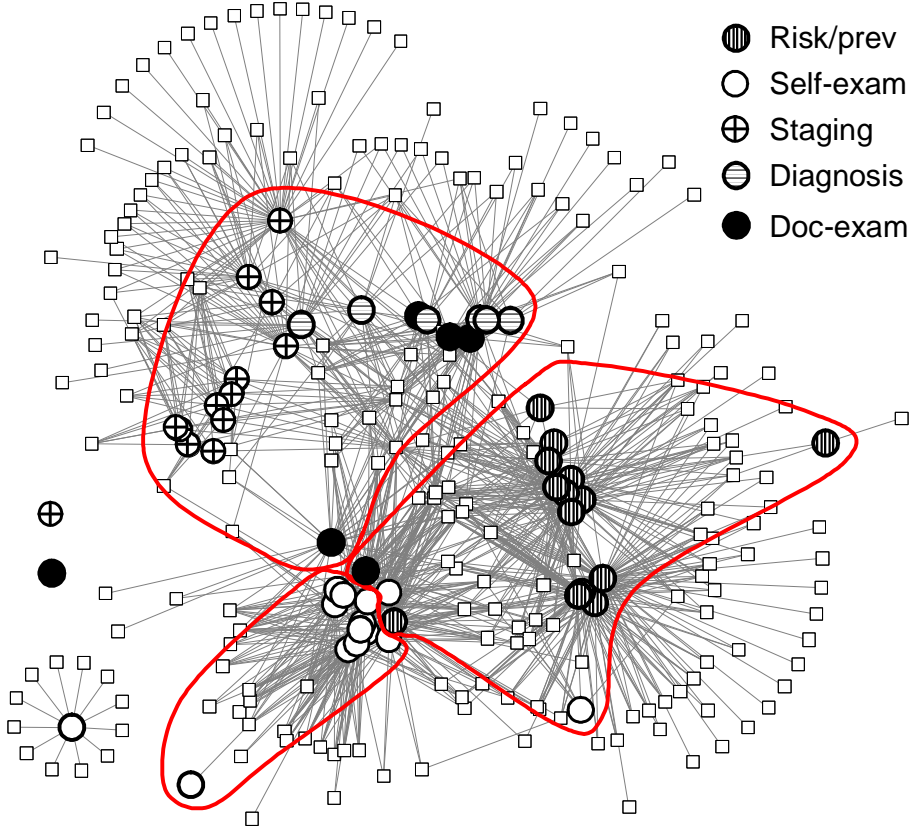}
\caption{Combined scatter networks for all five melanoma topics
(corresponding to fill of circles representing facts). The countours
indicate different communities of facts discovered by the community
finding algorithm discussed in
Section~\ref{community}.}\label{communities}
\end{figure}

Here we consider only which facts are connected to one another
through documents, rather than the documents and facts together.
This has to do with our conceptual question of reachability of
facts, and does not preclude applications where both may be
considered. As we can see in Figure~\ref{onemodediagram}, the
assortative network produces many isolated facts that occurred in
only a single document. The highly disassortative network on the
other hand had a single document that contained all the facts, thus
linking them together in a complete clique. Finally, the slightly
disassortative network produced two connected components.

For nearly all melanoma topics, the facts are located in a single
connected component. Furthermore, since some pages mentioning
different melanoma topics act as connectors, facts pertaining to all
five topics combine to form an even larger connected component shown
in Figure~\ref{communities}. The three isolated facts stem from
different topics. A stage calculator tool was not present on any of
the documents. Likewise, no documents mentioned the fact that during
a doctor's visit, the doctor may ask about the patient's history of
UV exposure. Finally, even though twelve documents listed resources
for locating a dermatologist, they did not mention any other
melanoma facts, and therefore that fact is isolated in the network.
While it is informative to know whether facts are connected through
some path, it is further important to know whether that path or
paths are short (the small world effect) and whether some facts are
clustered together, indicating that they may further form a cohesive
subtopic.

\subsubsection{The Small World effect: shortest paths and
proximity\label{smallworld}} The small world effect is named after
the observation that any two people in the world are linked through
a short chain of acquaintances~\cite{milgram67,wattssw}. As we saw
in Figure~\ref{onemodewlabels}, our networks become quite dense when
one takes a one mode projection. This means that many facts are
connected through one hop (a document that mentions them both), or
at most two. However, the chance of accessing exactly the one
document that mentions both facts may be small. Therefore,  in order
to effectively evaluate the proximity of two facts, we need to
consider more than just the shortest paths between them.

\begin{figure}[htbp]
\centering
\includegraphics[width=0.55\columnwidth]{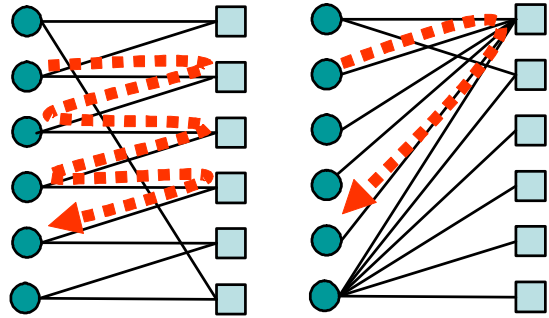}
\caption{ Two hypothetical networks with different average shortest
paths. (a) is a large world, since documents need to be traversed in
sequence for each additional fact reached (b) is a small world since
all facts are connected by a single
document}\label{shortestpathdiagram}
\end{figure}

\begin{figure}[htbp]
\centering
\includegraphics[width=0.85\columnwidth]{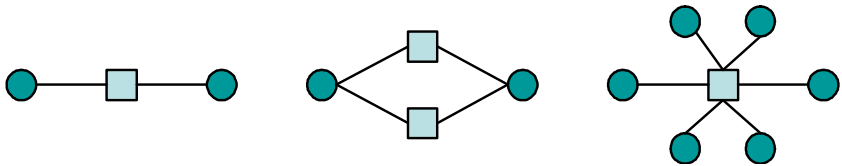}
\caption{Different configurations of how two facts (circles) may be
linked together by webpages (squares)}\label{proximitydiagram}
\end{figure}

Figure~\ref{proximitydiagram} shows three of the many different
configurations of how two facts may be linked: (a) two facts are
mentioned together only in one document, and that document contains
no other facts, (b) the facts co-occur in two documents, and (c) two
facts co-occur in only one document, and that document contains many
other facts. Intuitively, the tie between the facts is stronger in
(b) than in (a) because the facts are linked together by multiple
pages. The tie is also stronger in (a) than in (c), because the
document in (c) is a general document containing many facts, so the
co-occurrence of any of those facts is not as significant. A measure
that captures proximity in this way is the cycle-free effective
conductance (CFEC)~\cite{koren06proximity}. It represents the
probability that a user traversing the graph randomly will reach the
second fact before looping back to a previously visited fact or
webpage. Naturally, the more documents bridge the two facts, the
more likely a user is to reach the second from the first.

In the melanoma network, facts within a topic are on average more
proximate (CFEC = 1.67) than facts between topics (CFEC=0.35). The
risk topic facts are overall most central in the network, having a
proximity of 0.89, with the fact that high UV exposure is a risk
factor for melanoma occupying the most proximate position to all
other facts in the network. The other four topics had average whole
network proximities of 0.40-0.43. In the following sections we will
examine in more detail how one can characterize the high overlap in
facts between sets of documents and apply community finding
algorithms to discover them automatically.

\subsubsection{Clustering\label{clustering}} It is often useful to
compute a network's clustering coefficient, to measure how much
local structure there is in the network relative to a random graph.
The high clustering observed in many real-world networks reflects
the prevalence of closed triads~\cite{newman03structure}. This
indicates that in most real networks, if vertex A is connected with
vertices B and C, then the probability for B and C to be connected
is higher than expected at random. Since by definition closed triads
do not exist in a bipartite network, here we define clustering in
terms of cycles of length four. For example, we may ask: if a fact A
is contained in documents X and Y, and B is also contained in
document X, then what is the probability for B to be contained in
document Y as well? Similarly, if document X has facts A and B, and
document Y also has fact A, then what is the probability that
document Y also has B? The answers to these questions could help us
to understand the overlap of facts in documents. First, we give a
definition of the clustering coefficient in a bipartite graph with
$m$ documents and $n$ facts:
\begin{equation}
C_4 = \frac{4\times \mathrm{number~of~cycles~of~length~}
4}{\mathrm{number~of~connected~quadruple~of~vertices}}
\end{equation}
We choose this definition in lieu of the vertex-centric definition
of the bipartite network clustering coefficient proposed by Lind et
al.~\cite{lind05bipartite}, because it more directly answers the
questions posed above.

In the actual bipartite
 graph constructed by using our melanoma-related data,
there are 28,831 cycles of length 4; and 316,073 connected
quadruples. Thus, the clustering coefficient is $C_4 =
\frac{4\times28831}{316073}=0.365$, which is significantly ($p <
10^{-12}$) larger  than in randomized, degree-preserving versions of
the same graph ($C_{4}=0.242\pm0.004$). From the above analysis on
the clustering coefficient, we can see that in information networks,
facts and webpages are more likely to be correlated and overlapped
than if they were randomly distributed. In the following section we
will take advantage of the clustering of facts to automatically
discover ``communities'' corresponding to different topics.

\subsubsection{Community Structure}\label{community} 
\begin{figure}[htbp]
\centering
\includegraphics[width=0.35\columnwidth]{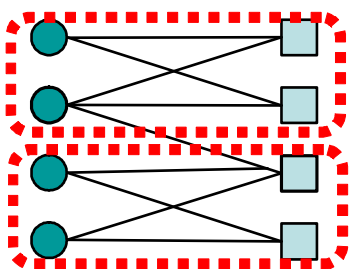}
\caption{An example of communities (groups of facts all shared by
many of the same documents) which have dense internal edges but only
a few edges between them.}\label{communitydiagram}
\end{figure}

As we saw in the
previous section and illustrate in Figure~\ref{communitydiagram},
facts may be ``clustered", many of the same facts co-occur in the
same set of documents. We may be interested in identifying such
clusters  because they may correspond to different subtopics. For
small networks, this question may be answered through a visual
analysis. For example, Figure~\ref{risknet} shows two clusters of
co-occurring facts within the risk/prevention network corresponding
to two subtopics: risk (e.g. {\em Weakened Immune System, Family
History}) at the top and prevention (e.g. {\em Sun Screen, Avoid
UV}) at the bottom. However, larger networks such as the inter-topic
scatter network shown in Figure~\ref{communities}, often need to be
analyzed using a community-finding algorithm. The algorithm by
Clauset et al.~\cite{clauset04community} aggregates nodes into
groups (communities) and stops when it has achieved maximum
modularity, meaning that the number of edges within communities
compared to the number of edges between communities is much higher
than if they had been randomly arranged. Although this algorithm has
so far been applied to one mode networks, we find that simply
applying it to a bipartite network without modification produces
intuitive results: facts that co-occur in many documents are placed,
along with those documents, in the same community.

If the five topics had mostly dedicated pages for each topic, then
the algorithm would identify 5 communities in the connected part of
the graph. However, as marked with the red borders in
Figure~\ref{communities}, the algorithm found only three communities
consisting of facts and pages from: (1) doctor's exam, diagnostic
tests, and disease stages, (2) self-exam, and (3) risk/prevention.
The algorithm revealed that the three topics in the first community
are strongly related through their co-occurrence on pages, and
indeed, all three topics deal with diagnosis. In contrast, the other
two communities tend to have more dedicated pages. Besides lumping
three of the topics together, the algorithm also assigned three
facts to a topic different from the one assigned by human experts.
Two self-exam facts, {\em checking the entire skin surface} and {\em
checking for irregular moles} (shown as solid white circles in the
right-hand-side fact group in Figure~\ref{communities}) are placed
in the community with risk and prevention facts. This probably
occurs because self-exams are a basic part of risk and prevention.
Similarly, one basic fact about doctor's exams, that one should have
them regularly (shown as a solid black circle) is also grouped under
risk and prevention, probably because a doctor's exam is also a very
basic part of risk and prevention. Thus we observe that community
finding connects very general facts, which may otherwise be grouped
into their more specific topics, with other general and important
facts. In the next section we will examine how network centrality
measures can help us identify such facts that bridge different
topics.

\subsubsection{Centrality and betweenness}\label{betweenness} Given
that information tends to be highly scattered and users may begin
their exploration on an arbitrary page, we would like to determine
the centrality of each page with respect to other facts and
documents in the network. A user who first lands on a central page
has a greater chance of rapidly learning the facts, and this has
important implications e.g. for search engine ranking. Here we
consider two intuitive notions of centrality. The first is simply
the ``degree centrality" of the vertex, namely the number of facts
contained in the webpage. A webpage with many facts will be central
because it will be associated with many other webpages through all
the facts it contains. Naturally, a user will find more
comprehensive information in a document with high degree than one
with lower degree.

\begin{figure}[htbp]
\centering
\includegraphics[width=0.65\columnwidth]{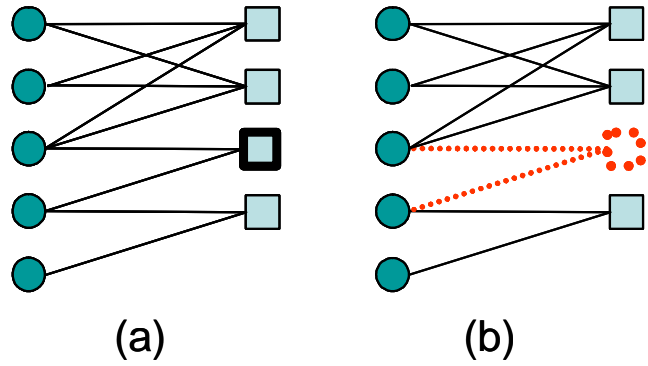}
\caption{The document outlined in bold has the highest betweenness,
and would actually disconnect a part of the network if
removed.}\label{betweendiagram}
\end{figure}

\begin{figure}
\centering
\includegraphics[width=0.75\columnwidth,angle=270]{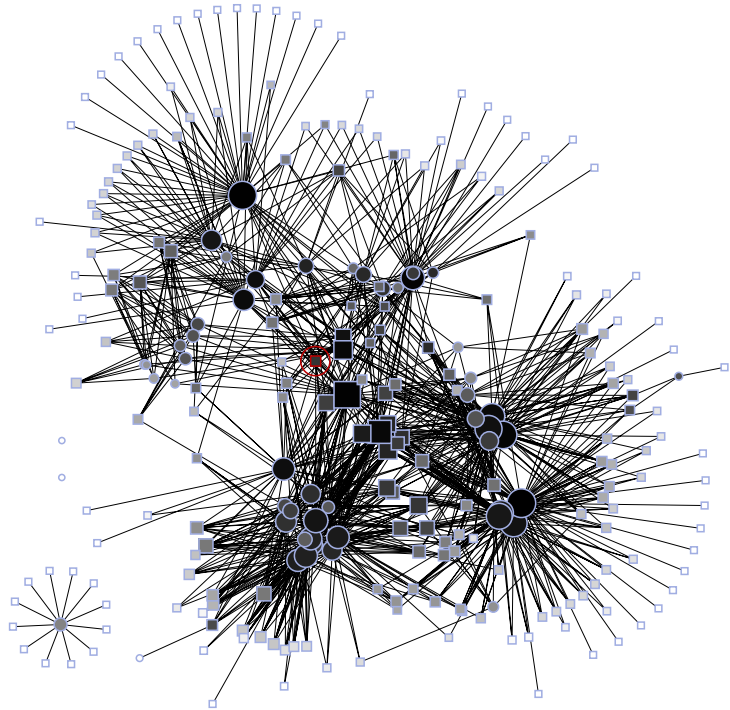}
\caption{The same scatter network as in Figure~\ref{risknet} with
nodes sized according to degree and shaded according to betweenness.
The circled document is a skincancer.org page with low degree but
high betweenness.}
\end{figure}

Having a lot of facts, however, does not necessarily guarantee broad
coverage. As we saw in Section~\ref{community}, groups of pages tend
to specialize in specific subtopics. It is therefore possible for a
page to contain many facts, but for those facts to all be on the
same subtopic. A user accessing such a page may or may not become
aware of other subtopics that could guide their further search.
Because of this, we consider a second centrality measure, the {\em
betweenness centrality}~\cite{freeman77centrality}. Betweenness
measures not how many neighbors a vertex has (though that frequently
correlates very highly with its betweenness), but rather how many
shortest paths pass through that vertex. The formula for computing
betweenness is:
\begin{equation}
C_B(n_i) = \sum_{j<k} g_{jk}(n_i)/g_{jk}
\end{equation}
Where $g_{jk}$ is the number of shortest paths connecting vertices
$j$ and $k$, and $g_{jk}(n_i)$ is the number of paths that vertex
$i$ is on. Figure~\ref{betweendiagram} illustrates the concept of
betweenness on a simple scatter network. The third document from the
top contains two facts, and so has slightly below average degree
centrality, but it happens to link facts that do not otherwise
co-occur. Even though removing the document leaves every fact still
in at least one of the documents in the network, the bottom two
facts may now not be discovered by a user since they do not co-occur
with any of the facts in the rest of the network.

While, in principle, pages with high betweenness could be fact-rich
or fact-poor, our results show that in the inter-topic melanoma
network, the node degree is highly correlated with node betweenness
($\rho=0.76$). This means that it is typically the fact-rich
webpages that bridge the different topics. But it is at the points
where the two measures diverge that we can glean interesting
insights about the scatter networks. For example, the page {\em
melanomanet/medical\_diagnosis.htm} on the website {\em
www.skincarephysicians.com} corresponds to a page on diagnosis with
a modest number of facts (7). But since these facts span all 5
topics, the page has very high betweenness. A user discovering this
page while searching for information on diagnosing melanoma would
have the opportunity to be exposed to other topics such as doctor's
visits and melanoma staging. Similarly, the fact that a skin biopsy
is the only way to be certain if a mole is melanoma is mentioned in
only 9 pages, but these pages themselves are scattered across
different topics, so that this rare fact becomes a central player,
linking diagnosis, self-exams, and doctor's visits as topics.
Naturally, documents that link to rare facts are assigned higher
betweenness because they present one of the few ways to reach those
facts. In the following section we will test to what extent these
high betweenness pages are essential in binding the scatter network
together.

\subsection{Robustness\label{robustness}}
In the context of information scatter, robustness (also termed
resilience)~\cite{achilles_heel,Jeong2001} is important to consider
because it relates to how easily comprehensive information can be
found if some webpages are removed. This is not just a matter of
rare facts being contained in a few documents and therefore the
removal of those documents being damaging. Rather it is a question
of whether different facts about the same topic will still be
discoverable, if documents that contained them together are removed.
As in Section~\ref{connected}, we will consider the connected
components in the scatter network. Remember that within each
component any fact can be reached from any other by iteratively
traversing edges, corresponding to a user alternately reading about
new facts on a webpage and using those facts to search for
additional webpages. In order to address the question of which
documents are essential for comprehensive search by maintaining the
connectivity of the network, we systematically remove nodes
according to each of the following approaches in turn:
\begin{enumerate}
\item  {\bf Random removal}: A random fraction of the documents are
removed.
\item  {\bf Targeted removal of the most fact-rich documents}: A fraction of
documents are removed in order of decreasing degree.
\item  {\bf Targeted Removal of the documents with highest betweenness}: A fraction of documents are removed in order
of their betweenness in the network.
\item {\bf Website Removal}: All
documents hosted by a particular site are removed (simulating a
particular host being down).
\end{enumerate}

\begin{figure}[htb]
\centering
\includegraphics[width=0.40\columnwidth]{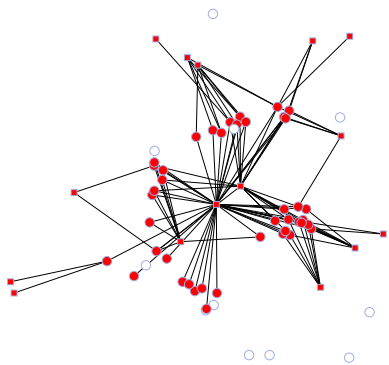}
\caption{The effect of randomly removing 90\% of the webpages from
the scatter network of melanoma facts and documents. Solid nodes
(both pages and facts) are in the giant connected component, the
white circle nodes are facts that are disconnected from it.}
\end{figure}

The melanoma scatter network proves to be quite resilient to random
node removal. Fifty percent of the documents may be removed with no
or only one fact becoming detached from the network. Even when {\em
ninety} percent of the documents are missing, most of the facts stay
connected through a few surviving fact-rich pages. The most
vulnerable subtopic is that of determining the stage of a melanoma
case, since there are few documents containing those facts, and if
they fall among those that are removed, all the facts in the
subtopic may be disconnected not only from the rest of the network,
but from each other as well.

\begin{figure}[htb]
\begin{center}
\begin{tabular}{lr}
  \includegraphics[width=0.45\columnwidth,angle=270]{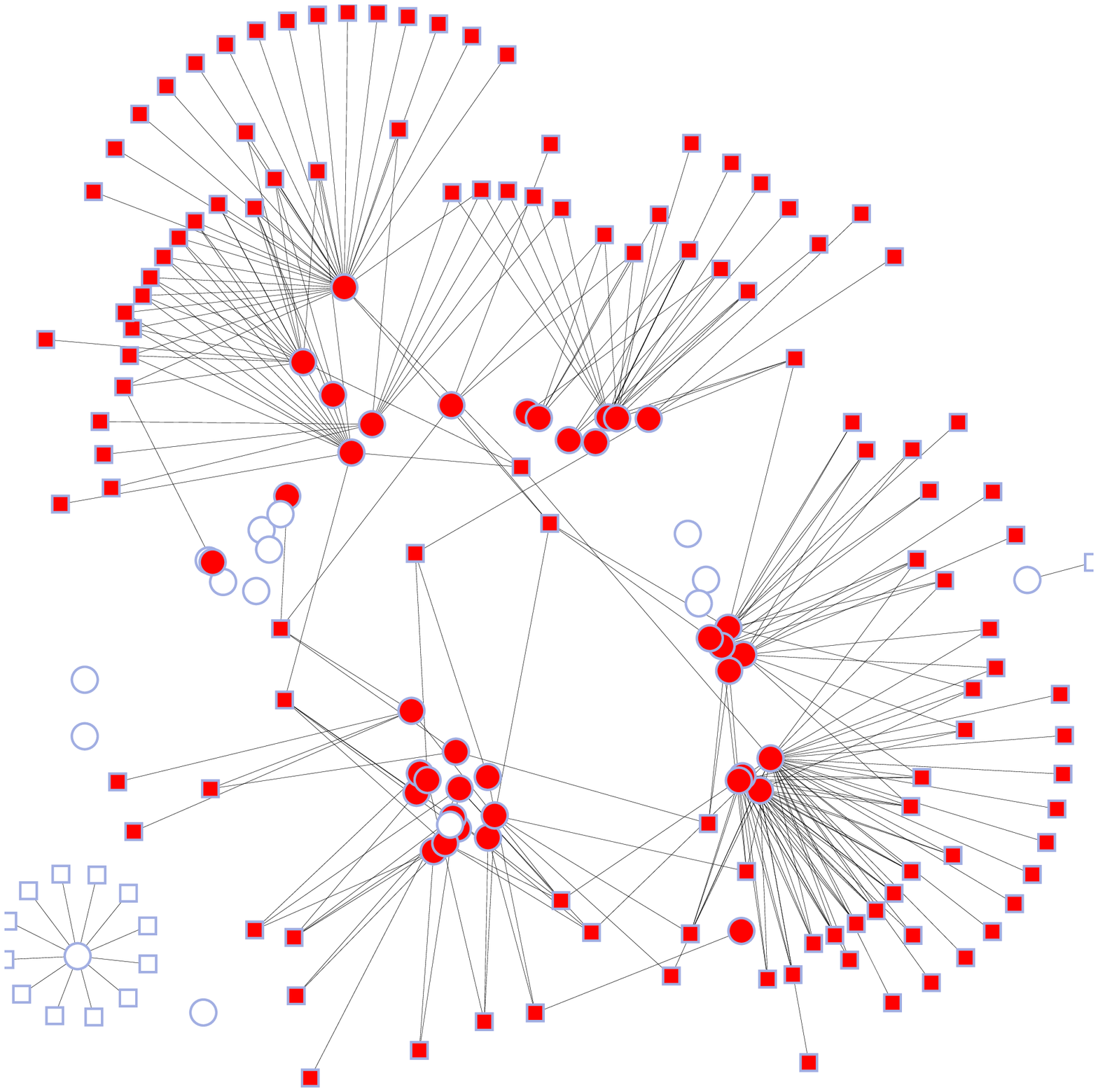} &
  \includegraphics[width=0.45\columnwidth,angle=270]{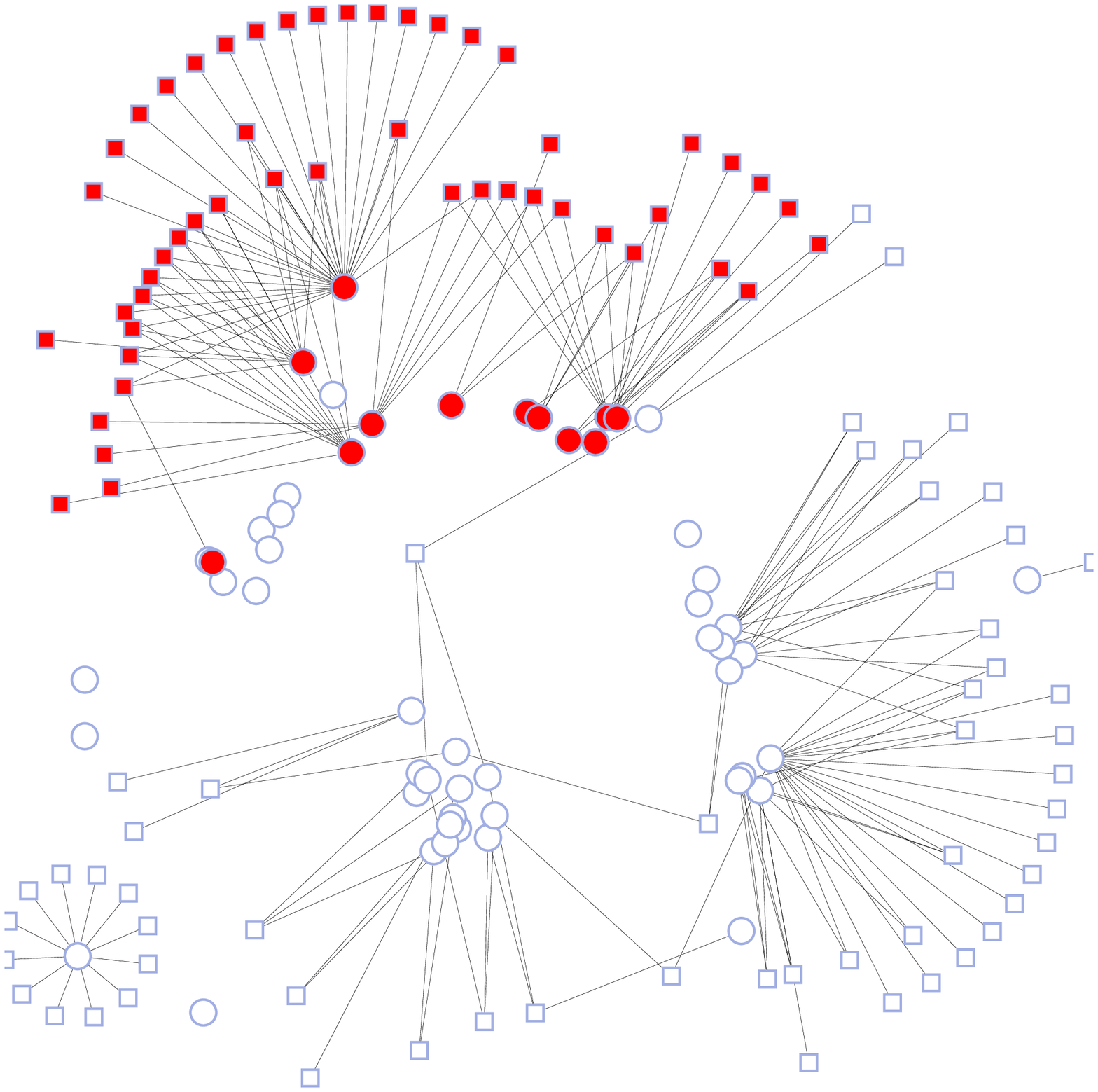}\\
  (a) 40\% webpages removed& (b) 50\% webpages removed\\
\end{tabular}
\caption{The effect of the removal of the fact-richest pages on the
connectivity of the network.} \label{degremoval}
\end{center}
\end{figure}
\begin{figure}[htbp]
\begin{center}
\begin{tabular}{lcr}
  \includegraphics[width=0.46\textwidth,angle=270]{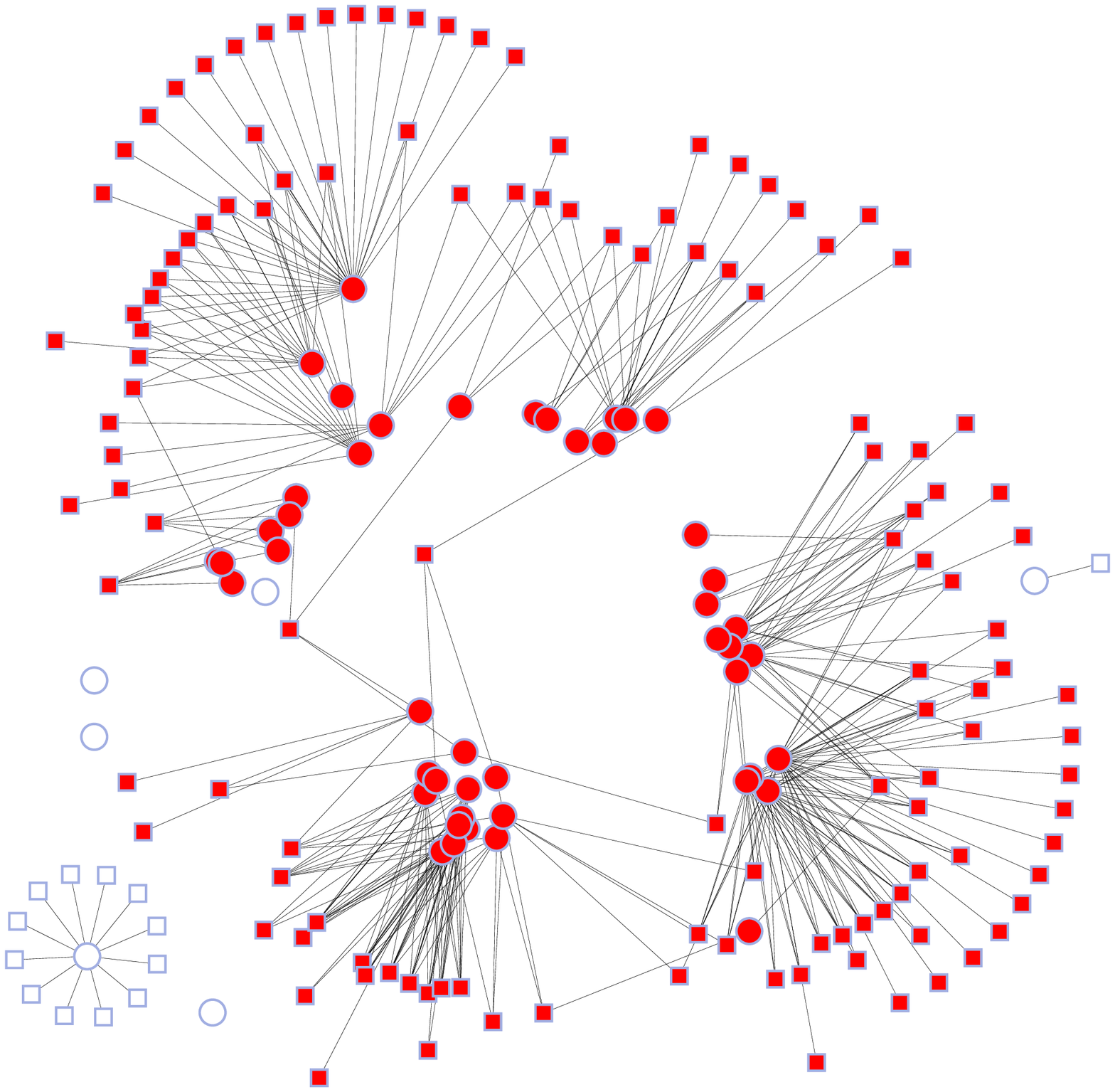} &
  \includegraphics[width=0.46\textwidth,angle=270]{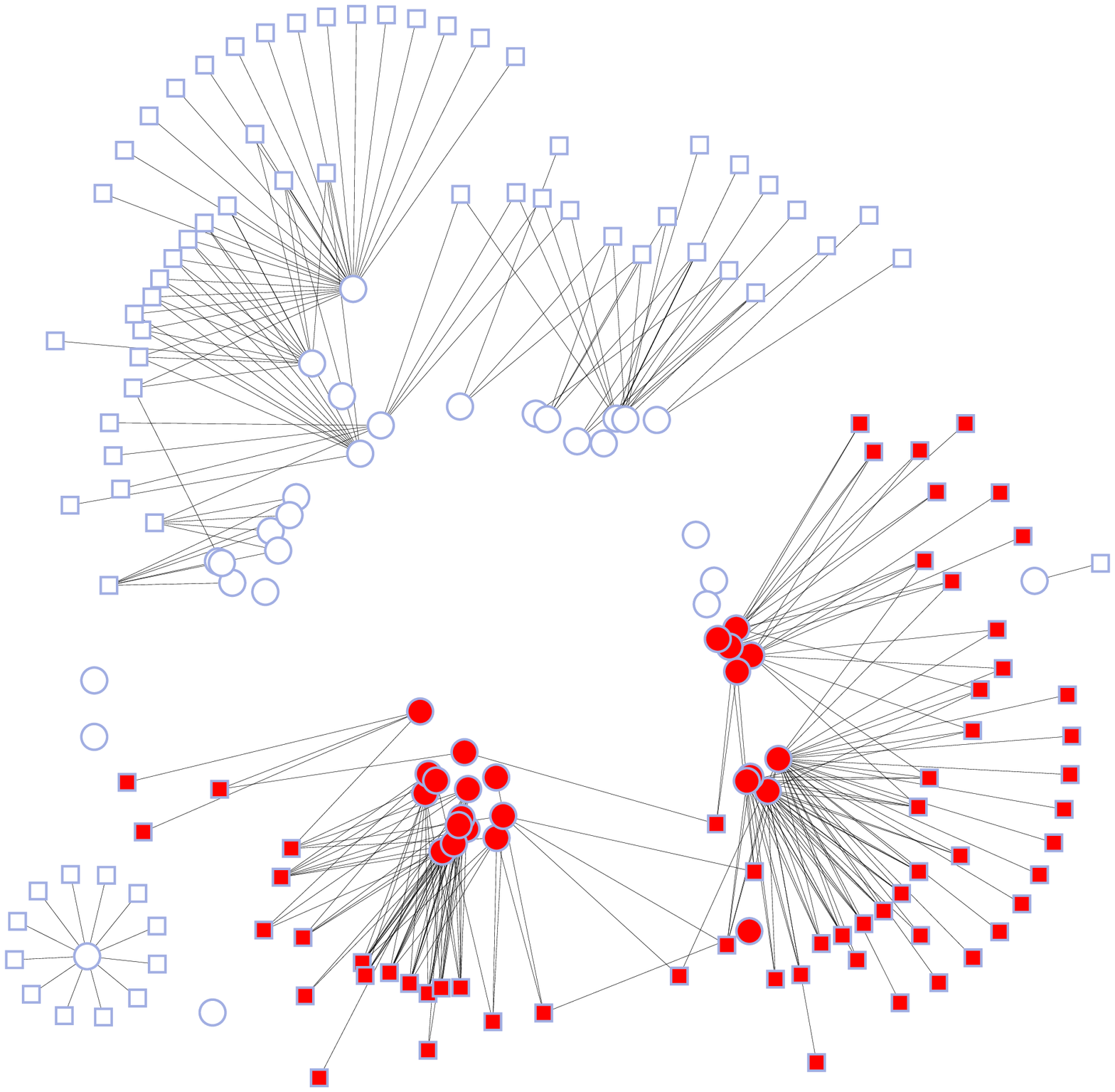}
  \\
  (a) 35\% webpages removed& (b) 40\% webpages removed\\
\end{tabular}
\caption{The effect of the removal of pages with highest betweenness
on the connectivity of the network.} \label{betremoval}
\end{center}
\end{figure}

As shown in Figure~\ref{degremoval}, targeted removal of the most
fact-rich pages is obviously much more damaging than random removal.
Whereas random removal required 90\% of the documents to be removed
in order to start disconnecting facts, this starts to occur after
removing the top 40\% of most fact-rich documents. By the time 50\%
of the pages are removed, most of the facts are disconnected from
one another. In Section~\ref{betweenness}, we discussed how pages
with high betweenness, in addition to usually mentioning many facts,
also bridge different topics by including disparate facts that are
not mentioned together in many or any other documents. Intuitively,
one might expect that removing such bridges would have an even
greater effect on the network than simply removing pages with many
facts~\cite{holme02attack}. This is confirmed in
Figure~\ref{betremoval}, which shows that facts start to become
disconnected after approximately 34\% of documents with the highest
betweenness are removed. Most facts become disconnected after 40\%
of the pages are removed. Comparing this result with the targeted
removal in the order of nodes' degree, we can see that documents
with higher betweenness are more important in the connectivity of
the graph than ones with only more facts. This tells us that
betweenness is more important to comprehensive searching. Referring
back to the toy network in Figure~\ref{betweendiagram}, we note that
the top two documents have the highest number of facts, but removing
either one of them would not disconnect the network since their
facts are replicated elsewhere.  However, it is the bottom two
documents, containing two facts a piece, that play a more important
role of connecting the graph.

Finally, we simulate what would occur if one of the websites hosting
the documents would disappear. We find that no removal of a single
website causes any of the facts to become disconnected from the
network. The one exception is that the fact that one should use body
maps to mark the location of existing moles is found only on the
site \texttt{melanoma.com}. All other facts are mentioned by
documents hosted by at least two different sites. This is an
encouraging result, for it means that if a site were to become
unavailable temporarily due to a server problem or permanently due
to an intentional decision, comprehensive information about the
topic of melanoma would still be available online.

In conclusion, the melanoma scatter network is remarkably robust,
showing much redundancy in fact coverage that makes the network
resilient to both random failure and an intentional attack (which is
unlikely to occur for medical subjects such as melanoma but may be
an issue for more controversial topics subject to censorship). It
would be interesting to examine in future work the robustness of
other, sparser, networks corresponding to topics where facts are
even more scattered.

\section{Combining fact-page and hyperlink networks}\label{hyperlink}
So far we have only considered the scatter network consisting of
pages and the facts they contain. We have argued that such a network
may model a search process wherein a user discovers a new fact on
one page and subsequently discovers other pages by searching for
that one fact. However, the description would not be complete
without considering how these pages are connected together via
hyperlinks. Here we present only a preliminary analysis of the
intersection of the scatter and hyperlink networks with the aim of
answering the question of how complete the fact coverage is of both
a random and smart web surfer. Because the webpages in the melanoma
scatter data set described above were not cached electronically, we
were unable to extract the hyperlinks between them. Instead, we
repeated the procedure described in Section~\ref{datagathermethod},
just for queries relating to melanoma risk and prevention, and
downloaded all pages matching those queries from 20 high quality
sites. We then automatically extracted all the hyperlinks between
those pages. Very few hyperlinks crossed different websites, and
even fewer were deep links to specific content on melanoma risk and
prevention. We therefore focused on links within just two websites,
\texttt{wikipedia.org} and \texttt{melanoma.com}, each of which
contained both many facts and a rich hyperlink structure between
documents matching the query. The presence or absence of each of the
14 melanoma risk and prevention facts was manually determined.

\begin{figure}
\centering
\includegraphics[width=0.75\columnwidth,angle=0]{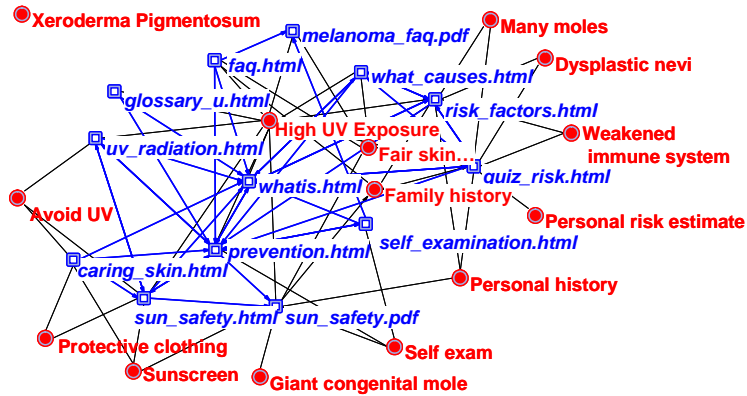}
\caption{The scatter and hyperlink network of risk and prevention
facts on \texttt{melanoma.com}. Blue squares are pages, blue thick
lines are hyperlinks, red circles are facts, and thin black edges
indicate the presence of a fact in a
page.\label{melanomacomhyperlink}}
\end{figure}

\begin{figure}[htb]
\centering
\includegraphics[width=0.75\columnwidth,angle=0]{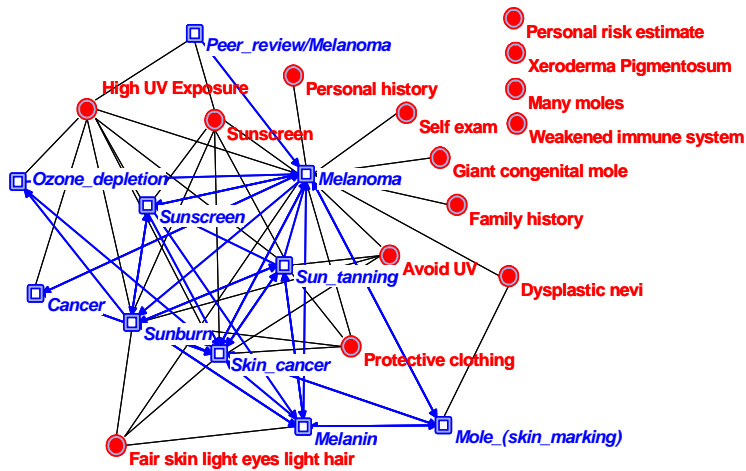}
\caption{The scatter and hyperlink network of risk and prevention
facts on Wikipedia.\label{wikipediahyperlink}}
\end{figure}

We visualized the overlapping network, showing directed hyperlinks
between webpages containing at least one fact, and all the edges
between the webpages and the facts they contain.
Figure~\ref{melanomacomhyperlink} shows the \texttt{melanoma.com}
website, containing 36 matching pages, 14 of which contained at
least one fact individually and 13 facts collectively. It is
immediately apparent that the risk-focused pages tend to both share
many facts and link directly to one another. The same is true of
prevention pages, which link to one another and contain facts about
how to protect against exposure to UV radiation. Finally, the two
sets of webpages are themselves bridged by a smaller number of
hyperlinks and by indirect links from general pages about melanoma.
Our second example, shown in Figure~\ref{wikipediahyperlink} is that
of the website \texttt{en.wikipedia.org}. Here we do not see a clear
grouping of risk vs. prevention pages. Rather there is a fact-rich
central ``melanoma'' page and closely linked pages on related
medical conditions and sun exposure topics. There were 64 Wikipedia
pages matching at least one of the queries, 13 of which were
non-duplicate pages containing at least one fact. Altogether 10 of
the 14 facts were mentioned in this set of heavily-interlinked
Wikipedia pages.

\begin{figure}[htb]
\centering
\includegraphics[width=0.75\columnwidth,angle=0]{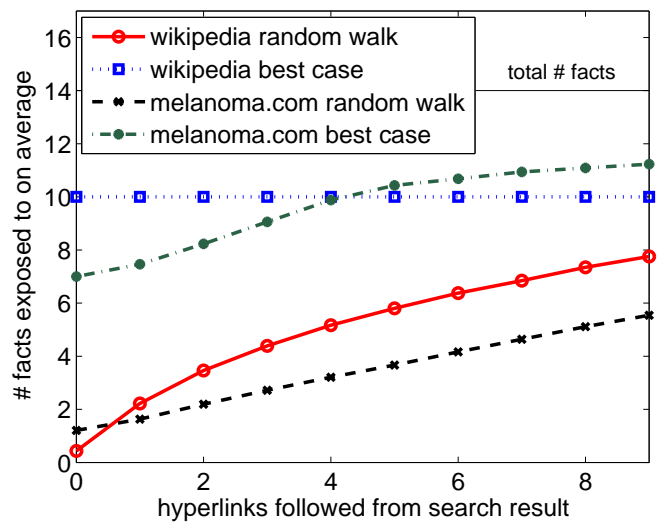}
\caption{Speed with which users can locate facts by navigating
hyperlinks on two sites.\label{exploration}}
\end{figure}

Finally we are ready to evaluate how rapidly a user will read all
the facts if they follow hyperlinks. We consider two conditions:
\begin{itemize}
\item {\em random walker}: The user randomly lands on any one of the pages
matching the queries within the website. He/she then continues to
follow hyperlinks (restricted to the subset of matching webpages),
avoiding backtracking to an already visited page, and jumping to
another random search result if stuck.

\item {\em smart surfer}: The user lands on the most fact-rich page (the
``melanoma" page on \texttt{wikipedia.org} and the
``risk\_factors.html" page on \texttt{melanoma.com}, which happened
to be the \# 1 or 2 results on both MSN search and Google search.
From the best page, the user follows hyperlinks, but only those that
point to pages containing at least one fact (we are assuming the web
surfer can somehow sense which pages will contain facts). If they
get stuck, they jump back to the 'best' page and start again.
\end{itemize}

Figure~\ref{exploration} shows how many facts are discovered on
average on both websites as a function of the number of hyperlinks
followed (the 0th hyperlink is the search engine result that was
initially selected). We observe that selecting the most fact-rich
search result makes a very significant difference. Strikingly,
visiting the main Wikipedia page about melanoma reveals all of the
available facts on the site. But even by performing a random walk
along hyperlinks starting at any Wikipedia page matching the query,
a user is able to slowly discover facts.  On the other hand, on
\texttt{melanoma.com}, a user continues to discover additional facts
even after the 10th step, eventually surpassing the number of facts
available on Wikipedia. On both sites, the rich hyperlink structure
allows users to explore the entire available fact space through
browsing. This is of course, only a preliminary analysis, one we
would like to repeat on a larger scale.

\section{Conclusions and future work}
In the above analysis we looked at information scatter through many
different interwoven network properties. They all pointed to the
importance of knowing how information comes together on the Web, and
not just counting the number of facts in each page. We saw that
assortativity has important implications for the ease with which a
user can locate information simply by visiting pages with many
facts. Assortativity also influenced the connectivity of the
network, with disassortative networks (ones where a page with
multiple facts tends to contain rarer ones) having larger connected
components. Large connected components imply that a user could in
principle discover information in a comprehensive manner by
alternating between discovering new facts in pages and discovering
new pages by searching for new facts online. However, the distance
the user would need to traverse in this network is also important -
facts which are proximate will co-occur in many of the same
documents and so one will quickly be discovered along with the
other. High co-occurrence implies clustering - groups of facts that
appear together in many of the same webpages. We demonstrated that
these clusters of facts may be discovered by using community finding
algorithms. The automatically assigned groups closely correspond to
expert assigned relationships between facts and topics. But where
they diverge, we find that the ``misclassified" facts are either
bridging topics or are not ``misclassified" at all. We also examined
centrality of different documents in the topic space, and studied
the effect of removing central webpages in terms of whether users
would still be able to discover most of the facts through fact
co-occurrence. For most methods, such as community discovery, we
could use existing algorithms and metrics without modifications. For
others, such as degree correlations and clustering, we adapted the
definitions to fit the context.

In future work, we intend to extend our analysis to a range of
domains by automating the extraction of keywords from pages and
identifying which keywords occur in which pages. Furthermore we
would like to apply our network analysis to improve search
algorithms for the purpose of assisting users in finding
comprehensive information. For example, scatter-hyperlink networks
(of which examples are shown in Figures \ref{melanomacomhyperlink}
and \ref{wikipediahyperlink}) provide a powerful representation for
the design of future comprehensive search engines. Unlike current
search engines which provide a ranked list of webpages but are
agnostic to the scatter of facts about the topic across the
webpages, future comprehensive search engines could provide a
portfolio of pages that directly address information scatter. This
portfolio of pages will not only collectively provide all the facts
about the topic, but will also provide a subset of active hyperlinks
on the pages which guide users to other pages which contain new
facts. The portfolio of linked pages therefore will leverage the
existing structure of links within a site and help users make sense
of the information in the context of existing webpages and their
link structure. The above search systems will take as input a
general topic (e.g. melanoma risk/prevention) and then use cluster
analysis techniques to identify common terms within relevant pages
as a proxy for facts. A scatter-hyperlink network of the returned
pages will then be constructed and analyzed to identify for instance
term-rich pages with many hyperlinks to other pages with terms not
contained in first page. Networks of facts and hyperlinks can
therefore be combined to help users in the future not only to find
highly scattered information, but also to rapidly make sense of it.

\section{Acknowledgments}
We would like to thank Dale Hunscher and Yan Qu for many insightful
discussions and Sameer Halai and Debbie Apsley for assistance in
collecting the data. We would also like to thank Yehuda Koren,
Stephen North, and Chris Volinsky for use of their network proximity
algorithm. This work was supported in part by an Accelerating Search
in Academic Research Grant from Microsoft.
\\
\bibliographystyle{plain}
\bibliography{../justinitials}
\end{document}